\documentclass[]{jpsj3} 
\usepackage{graphicx,color}
\usepackage{bm}

\def\({\left(}
\def\){\right)}
\def\aa{\textbf{\textit{a}}}
\def\kk{\textbf{\textit{k}}}
\def\pp{\textbf{\textit{p}}}
\def\ww{\textbf{\textit{w}}}
\def\qq{\textbf{\textit{q}}}
\def\UU{\textbf{\textit{U}}}
\def\QQ{\textbf{\textit{Q}}}
\def\RR{\textbf{\textit{R}}}
\def\GG{\textbf{\textit{G}}}
\def\rr{\textbf{\textit{r}}}
\def\e{\mathrm{e}}
\title{ Significance of Off-Center Rattling for Emerging Low-lying THz Modes in type-I Clathrates}

\author{Tsuneyoshi \textsc{Nakayama}$^{1}$\thanks{E-mail: tnaka@eng.hokudai.ac.jp} and Eiji  \textsc{Kaneshita}$^{2}$}
\inst{$^{1}$Hokkaido University, Department of Applied Physics, Kitaku, Sapporo 060-0826, Japan\\
$^{2}$Sendai National College of Technology, Aoba-ku, Sendai 989-3128}

\abst{We show that the distinct differences of low-lying THz-frequency dynamics between type-I clathrates with on-center and off-center guest ions naturally follow from a theoretical model taking into account essential features of the dynamics of rattling guest ions.
Our model analysis demonstrates the drastic change from the conventional dynamics shown by on-center systems to the peculiar dynamics of off-center systems in a unified manner.
We claim that glass-like plateau thermal conductivities observed for off-center systems stem from the flattening of acoustic phonon dispersion in the regime $|\kk|<|\GG|/4$.
The mechanism is applicable to other systems such as glasses or relaxers.
}

\kword{phonon dynamics, clathrates, thermoelectricity, thermal conductivity, symmetry breaking}

\begin{document}
\maketitle






\section{Introduction}

Type-I clathrate compounds with the chemical formula II$_8$III$_{16}$IV$_{30}$ have recently attracted a great deal of attention in connection with the search of efficient thermoelectric materials~\cite{Chr10}.
This is because, though these compounds are crystalline, they exhibit glass-like
phonon-thermal conductivities~\cite{Nol98, Kep00, Sal01, Ben04, Ume05}.
Diffraction experiments~\cite{Nol00, Ben05, Chr06} and extended x-ray absorption fine-structure (EXAFS) studies~\cite{Bau05,Jia08} on these clathrates have revealed that rattling guest ions in cages take either the on-center or the off-center positions depending on the relative size of the cage to the guest ion.
It is remarkable that clathrates  with ``off-center" rattling guest ions exhibit
glass-like low-lying modes in the THz-frequency region~\cite{Nak02},
which have been observed in specific heat measurements~\cite{Ume05, Xu10},
inelastic neutron scattering experiments~\cite{Lee07, Chr08},
and optic spectroscopy~\cite{Tak06,Sue10,Mor09, Kum10, Mor11}.

There are several theoretical works on the THz-frequency dynamics
of type-I clathrates~\cite{Mat09, Yam09, Eng09}.
To complement these works, we consider in this paper another physical model
to investigate the differences between the on-center system (ONS) and the off-center system (OFS).
Our approach emphasizes the difference of ONS
 and OFS on the emergence of peculiar low-lying THz-frequency modes observed in type-I clathrates with
off-center rattling guest ions.
In addition,
we discuss the effect of the long-range electrostatic interaction between the rattling guest ions, which has been overlooked in other work.
Recent experimental data in terms of far infrared spectroscopy at low temperatures
by Mori $et~al.$~\cite{Mor11} for OFS provide important information on the difference about THz frequency dynamics from ONS.

We construct the equations of motion to determine the dispersion relations of acoustic and low-lying optic modes for ONS and OFS of clathrates.
The calculated results quantitatively clarify the distinct differences of the THz-frequency dynamics between ONS and OFS in the following points:  dispersion relations of acoustic phonons, behaviors of low-lying optic modes and characteristics of thermal transports driven by acoustic phonons.

\section{Description of  the Model}
We illustrate compounds with the chemical formula II$_8$III$_{16}$IV$_{30}$
as a prototype of type-I clathrates.
Rattling guest ions in these compounds  take the on- or off-center position
in cages of 14 hedron depending on the ratio of cage size and ionic radius of rattling guest ion.
For example, the rattling guest ions in 14 hedron in $\beta$-Ba$_8$Ga$_{16}$Sn$_{30}$ ($\beta$-BGS)
takes the off-center position at $U_0=0.43$~{\AA} from the center
of the tetradecahedral cage~\cite{Ume05}.
The deviation $U_0=0.43$ {\AA} is 7.4\% of
the nearest-neighbor distance $d=5.84$~{\AA} between Ba$^{2+}$ ions.
The shape of cages restricts
the motion of the off-center guest ions  into
quasi two-dimensional plane parallel to the hexagonal faces of the cage.
Hence, the potential energy for off-center guest ions is shaped like the bottom of a wine bottle
with fourfold hollows~\cite{Sal01, Mad05}.
On the other hand, the rattling guest ions in $n$-type $\beta$-Ba$_8$Ga$_{16}$Ge$_{30}$ ($\beta$-BGG)
take almost on-center positions in cages~\cite{Sal01,Chr06}
forming a potential shaped like the bottom of a ``wine glass".

We focus our attention on the lowest two bands (acoustic and optic modes)
in the THz-frequency range.
This is because these two modes are crucial for elucidating the mechanism controlling the differences.
We treat the cage as having mass $M$ and an effective charge $e_C^*$; and the rattling guest ion, mass $m$ and charge $e_G^*$.
We express the position of the $\ell$-th cage at time $t$
as $\RR_\ell+\RR_\ell(t)$,
where $\RR_\ell$ is the equilibrium position of the $\ell$-th cage center.
The  vector $\rr_\ell(t)$ represents a small displacement from  $\rr_\ell$.
The position of the rattling guest ion from  $\rr_\ell$ is defined by the
vector $\UU_\ell+\UU_\ell(t)$, where $\UU_\ell$ is
the equilibrium position of the rattling guest ion from $\rr_\ell$,
and $\UU_\ell(t)$ is a small displacement from  $\UU_\ell$ at time $t$.
Note that $\UU_\ell\neq 0$ in the case of off-center guest ion,
while the on-center case becomes $\UU_\ell=0$.

Scaffold cages compose cubic lattice~\cite{Nol00, Ben05, Bau05,Chr06},
which is invariant under translation by any lattice vector.
Each cage is elastically connected with nearest neighbor cages.
We denote the coupling strength by the harmonic force constants $f_{\|}$ and $f_{\perp}$,
which represent longitudinal (dilation) and transverse (shear) modes.
With these quantities, the potential energy is expressed by
\begin{align}
\label{eq:harmonic}
\nonumber
V_1&=\sum_{\ell,\gamma}
\frac{f_{\gamma}}{2}\vert\rr_{\ell,\gamma}-\rr_{\ell+1,\gamma}\vert^2\\
&=2\sum_{\kk,\gamma}f_{\gamma}\vert \QQ_{\kk\gamma}\vert^2
\sin^2\left( \frac{\kk\cdot\aa}{2}\right),
\end{align}
where  $\aa$ is a lattice vector, and $\gamma$ denotes the species of three modes: $\|$, $\perp$, and $\perp'$.
The second line is obtained by the Fourier transformation:
\begin{equation}
\label{Fourier0}
   \rr_{\ell,\gamma}=\frac{1}{\sqrt{N}}\sum_{\kk}\QQ_{\kk\gamma}
   \e^{i\kk\cdot\rr_\ell}.
\end{equation}

The rattling guest ions are trapped in anharmonic potentials,
for which we keep up to the 4th order of small displacement in the expansion.
This choice enables us to treat
both of ONS and OFS in a unified manner.
The anharmonic potential is given, in terms of relative displacement $\ww_\ell=\UU_\ell(t)-\rr_\ell(t)$, by
\begin{eqnarray}
\label{eq:anharmonic}
V_2=\sum_\ell\left[\frac{\xi}{2}
\vert\UU_{\ell}+\ww_{\ell}(t)\vert^2
+
\frac{\eta}{4}\vert\UU_{\ell}+\ww_{\ell}(t)\vert^4\right] .
\end{eqnarray}
Note here that the anharmonic potential is expressed by the relative coordinate $\ww_\ell$.
This allows us to legitimately treat the THz frequency dynamics of a coupled system consisting of guest ions and networked cages beyond the adiabatic approximation.
In addition, it is possible to make the Fourier transformation for $\ww_\ell$ on the $\kk$-space since $\ww_\ell$ is independent of $\UU_\ell$.

The potential $V_2$ involves two types of anharmonic potentials depending on the values of parameters $\xi$ and $\eta$.
Depending on positive or negative $\xi$, Eq.~(\ref{eq:anharmonic}) represents wine-glass or wine-bottle type potential where $\eta$ is positive constant.
In addition to $V_2$, we should take account of the electrostatic interaction energy
$V_3$ between off-center ions; we shall discuss this later.

\section{Potential for Off-Centered Guest Ions}
So far, our discussion has been quite general.
We now restrict attention to the case of OFS,
for which we impose the condition
$\xi<0$ and $\eta>0$ in Eq. (\ref{eq:anharmonic}) for $V_2$.
This parameter set yields a wine-bottle type potential,
which takes its minimum
$V_{2,min}=-\xi^2/(4\eta)$ at $\vert\UU_\ell\vert^2=-\xi/\eta$.
The non-zero $\UU_\ell$ allows us to rewrite the potential
(\ref{eq:anharmonic}) as
\begin{eqnarray}
\label{eq:wine}
V_2=
\frac{\xi}{2}\sum_{\ell}\vert\UU_{\ell}+\ww_{\ell}(t)\vert^2\left[ 1-\frac{\vert\UU_{\ell}+\ww_{\ell}(t)\vert^2}{2\vert\UU_{\ell}\vert^2}\right] .
\end{eqnarray}
Here we define the fluctuation around the equilibrium position of off-center
guest ions by introducing complex number given as
\begin{eqnarray}
\label{eq:rotation}
W_\ell=\e^{i(\theta_\ell+\delta\theta_\ell(t)/\sqrt{2})}\left[ U_0+\frac{h_\ell(t)}{\sqrt{2}}\right] ,
\end{eqnarray}
where $W_\ell$ is a complex number representing the vector $\UU_\ell+\ww_\ell$.
From Eq.~(\ref{eq:rotation}), the deviation from the center of cage
becomes $U_0\e^{i\theta_\ell}$.
The equilibrium angles $\lbrace\theta_\ell\rbrace$
are oriented at random reflecting glass-like behaviors observed in specific heats and thermal conductivities for OFS~\cite{Nol98, Kep00, Sal01, Ben04, Ume05}.

The form of Eq.~(\ref{eq:rotation}) is analogous to that introduced in the local gauge theory in relevance with the Higgs mechanism, in which
a covariant derivative plays a role for providing with masses to dynamical variables  $h_\ell(t)$ and $\delta\theta_\ell(t)$.
We take a different approach to give physical implications for these dynamical variable
as explained below.

The random orientation of the vector  $\UU_\ell$ originates from four hollows of the anharmonic potential~\cite{Sal01, Mad05} in addition to inter-site dipole-dipole interactions between off-center guest ions~\cite{Nak08, Kan09}.
It has been confirmed~\cite{Sal01} that off-centered guest ions experience a hindering potential $V_h(\theta)$ with a fourfold inversion symmetry along the azimuthal direction. The first-principles calculations have shown $V_h$ to be $\simeq 20$ K for Sr$^{2+}$ guest ions in $\beta$-SGG~\cite{Mad05}.

The effect of long-range electrostatic force has been treated in our papers~\cite{Nak08, Kan09}
for the elucidation of glass-like behaviours of OFS at low temperatures.
Electric dipole moments of guest ions provide long-range interactions with  dipoles in other cages.
To make our argument clearer,
let us illustrate as an example the two coupled dipoles $\pp_1$ and $\pp_2$, whose potential is given by $V_{12}=V_h(\theta_1)+V_h(\theta_2)+V_{12}(\theta_1, \theta_2)$
with the dipole interaction $V_{12}=(e_G^*U_0)^2\cos(\theta_1-\theta_2)/ R_{12}^3$ with the distance $R_{12}$ between the dipoles at 1 and 2.
Two global minima (maxima) in $(\theta_1,\theta_2)$ configuration space appear at  $|\theta_1-\theta_2|=\pi$~$( 2\pi)$.

This argument can be straightforwardly extended to the case of multiple pairs providing many potential minima in the configuration space $\mathcal{P}=(\theta_1,\theta_2,\theta_3, \cdots)$, where the potential function is $V_{123\cdots}= \sum V_h(\theta_i)+\sum\tilde{V}_{ij}$.
This configuration acts as a new hindering potential in addition to the four-fold inversion symmetric potential $V_h$.
The key yielding glass-like behaviors is that the dipoles $\pp_\ell=e_G^*\UU_\ell$ in OFS constitute an equilateral-triangle structure among next-nearest neighbours dipoles
and easily generate a frustrated situation necessary to glass-like properties.
The energy-scale of dipole-dipole interaction is estimated to be of the order of a few 10~K in Ref.~\cite{Nak08} which is of the same order of the barrier height of $V_h$.

\begin{figure}[t]
\vspace*{10mm}
\begin{center}
\includegraphics[width = 0.2\linewidth]{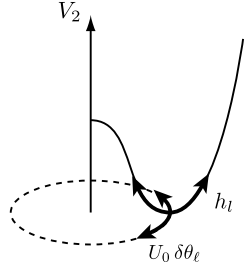}
\caption{Illustration showing the emergence of two freedoms of motion; stretching
$h_\ell(t)$ and libration $U_0\delta\theta_\ell(t)$.
}
\label{fig:Mode}
\end{center}
\end{figure}

This implies, in the temperature region $k_B T< V_h\approx $ a few 10~K, that off-center guest-ions execute angular fluctuation $\lbrace\delta\theta_\ell\rbrace$ from the bottom of the potential at $\lbrace\theta_\ell\rbrace$.
Thus, We treat the phase factor as a dynamical variable trapped in a hindering potential.
The variable $h_\ell(t)$ in Eq. (\ref{eq:rotation}) represents small fluctuation
along the radial direction directed at angle $\theta_\ell$.
Thus, we have a small displacement of rattling guest ion at $\ell$-th site to the first order
\begin{eqnarray}
\label{eq:rotation2}
w_\ell(t)=\frac{1}{\sqrt{2}}[h_\ell(t) +iU_0\delta\theta_\ell(t)]\e^{i\theta_\ell}.
\end{eqnarray}
We have depicted the situation in Fig.~\ref{fig:Mode}.

The equation (\ref{eq:wine}) becomes, under the situation $|\UU_\ell |>|\ww_\ell| $ using Eq. (\ref{eq:rotation})
\begin{eqnarray}
\label{eq:wine2}
   V_2=\frac{1}{2}\sum_{\ell}(\tilde{\xi}_s h_\ell^2+\tilde{\xi}_\theta
         U_0^2\delta\theta_\ell^2),
\end{eqnarray}
with the definition of the effective force constants for stretching and libration modes given below
\begin{equation}
\label{Effective}
\tilde{\xi}_s=-2\xi_s(1+\left\langle h_\ell^2\right\rangle /4U_{0}^2),~~
\tilde{\xi}_\theta=-2\xi_\theta(1+\left\langle \delta\theta_\ell^2\right\rangle /4).
\end{equation}
The effective force constants $\tilde{\xi}_s$ and $\tilde{\xi}_\theta$ in Eq.~(\ref{eq:wine2}) are introduced to redefine $\xi$ in Eq.~(\ref{eq:wine}) in order to express the differences between angular
motion and stretching motion.
It should be emphasized here that the effect of random orientation $\left\lbrace U_\ell\right\rbrace$ on effective force constants shall be involved in  newly introduced  force constant $\bar{\xi}_\gamma$ in Eq.~(\ref{eq:Fourier_anharmonic5}).

We have employed the thermal average in these expressions.
Thermal-averaged squared-displacements, $\langle h_\ell^2\rangle$ and  $\langle\delta\theta_\ell^2\rangle$, are proportional to temperature $T$.
The Raman scattering~\cite{Tak06, Sue10, Kum10} and far infrared spectroscopy~\cite{Mor09, Mor11} experiments show  that spectra monotonically increase with temperature $T$.
These observations indicate the relevance of anharmonic potential
via $\langle h_\ell^2\rangle$ and  $\langle\delta\theta_\ell^2\rangle$ since
$\tilde{\xi}_s$ and $\tilde{\xi}_\theta$ are directly related with the spectra as will be given in
Eq. (\ref{eq:optic_eigenfrequency}).

It should be emphasized that anharmonic terms in Eq.~(\ref{eq:wine2}) involved in $\tilde{\xi}_s$ and $\tilde{\xi}_\theta$ become irrelevant below a few 10~K.
Thus, we do not take into account the effect of anharmonic terms in the low temperature region discussed in this paper.
In this connection, we remark that, at sufficiently high temperatures above a few 10K, guest ions  execute free-circular motion and should recover the translational symmetry.

\section{Equations of Motion}
\subsection{Coupling between Acoustic Phonons and Guest Ions for Off-Center System}
Let us discuss the effects of the random orientation
of the equilibrium angles $\lbrace\theta_\ell\rbrace$ of off-centered guest-ions at the site $\ell$.
There are two aspects on the effects to the THz frequency dynamics.

The first is the coupling mechanisms between acoustic phonon modes and off-centered guest-ions trapped in cages.
There are two possible types of the coupling: the symmetry-sensitive coupling,
which is related to deformation of cages, and the symmetry-insensitive type related to the displacement of cages and guest ions.

The potential function Eq.~(\ref{eq:wine}) for off-centered guest ions manifests
the local potential interacting with cages.
The situation is analogous to localized electron-phonon coupling in insulators.
From the arguments of symmetry of modes for this case,
Raman-active stretching mode couples with longitudinal acoustic phonons accompanied with ``dilatation", and the infrared active libration mode does with transverse acoustic modes without dilatation.
Thus, when the symmetry-sensitive coupling is valid, the infrared active libration motion $U_0\delta\theta_\ell$ primarily concerns
with the transverse modes ($\perp$) without dilatation, while the radial stretching $h_\ell(t)$ does with the longitudinal mode ($\|$) with dilatation.

\begin{figure}[t]
\vspace*{10mm}
\begin{center}
\includegraphics[width = 0.4\linewidth]{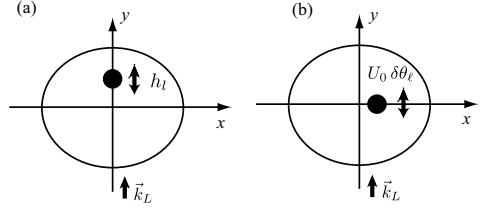}
\caption{Illustration showing the symmetry-insensitive coupling between acoustic phonons and guest ions. The figure (a) is the case that the off-center guest ion  takes the far-side position for incident longitudinal acoustic phonons with the wave vector $\kk_L$, and (b) shows the guest ion sits perpendicular position to the same longitudinal acoustic phonons.
The coupling becomes effective to parallel components of relevant displacements regardless of the transverse or longitudinal modes.
}
\label{fig:Coupling}
\end{center}
\end{figure}

The symmetry-insensitive coupling is independent of the symmetry of modes.
This coupling becomes relevant to parallel components of each displacement regardless of the transverse or longitudinal modes.
The situations illustrated in Fig.~\ref{fig:Coupling}(a) and (b) provide the implication of this coupling mechanism.
For example, consider the case that longitudinal acoustic phonons with polarization vector parallel to the $y$-axis are incident along $y$-axis when the guest ion takes the far-side position as depicted in Fig.~\ref{fig:Coupling}(a).
These longitudinal acoustic phonons dominantly couple with the stretching motion $h_\ell(t)$
parallel to the polarization vector.
Provided that the guest ion sits the right side along $x$-axis as shown in Fig.~\ref{fig:Coupling}(b), the longitudinal acoustic phonons incident along $y$-axis couple with the libration motion $U_0\delta\theta_\ell$ which is parallel to the polarization vector of the acoustic phonons.
Thus, the coupling constant continuously spans from the libration one $\tilde{\xi}_\theta$ to the stretching one $\tilde{\xi}_s$.
The same arguments hold for the coupling of transverse acoustic phonons.
These cause the broadening of the spectral width
through the distribution of force constants
 $\tilde{\xi}_\theta\leq\tilde{\xi}\leq\tilde{\xi}_s$.

Inelastic neutron scattering experiments are crucial to determine which coupling is dominant.
We describe this point in detail when showing the calculated phonon-dispersion relations
in Sec. V.

The second is on the line-width broadening of the spectra of optic modes.
Recent experimental data in terms of far infrared spectroscopy
at low temperatures~\cite{Mor11} for OFS provide interesting features on the line-width broadening.
They discovered that the spectral width of about 0.57~THz of the lowest-lying infrared active optic modes 0.71~THz at 7~K decreases with \textit{increasing} temperature.
This feature contradicts with the assumption that the anharmonicity of
the potential Eq.~(\ref{eq:anharmonic}) is a key element for interpreting the origin of the line-width broadening.

These experimental results~\cite{Mor11} suggest that the random configuration of  $\UU_\ell$ yields the broadening of optical spectra at low temperatures.
We claim that
the  estimated spectral width is given by $\bar{\Delta}\simeq z\mu_e^2/4\pi\epsilon_r\bar{r}^3$, where  $\bar{r}$ is the mean distance from off-centered guest ion to the nearest one, and $z$ is a number of the order of unity representing the effective coordination number dominated by a few dipoles which happen to be closest.


From the discussion made above, we reform the potential
function Eq.~(\ref{eq:wine}) as, under the situation $|\ww_\ell|<<|\UU_\ell|$,

\begin{align}
\label{eq:Reform}
\nonumber
   V_2&=\frac{1}{4}\sum_{\ell}\xi\vert\UU_\ell\vert^2\left[ 1-4\vert\frac{\ww_{\ell}}{\UU_\ell}\vert^2+ \mathrm{anharmonic~terms}\right]\\
   &=\frac{1}{2}\sum_{\ell,\gamma}\bar{\xi}_\gamma\vert\ww_{\ell,\gamma}\vert^2+\mathrm{const.}
\end{align}
It should be emphasized here that the newly defined effective force constant $\bar{\xi}_\gamma$ involves the effect of random orientation of $\left\lbrace U_\ell \right\rbrace$ in a manner as explained after Eq.~(\ref{eq:Fourier_anharmonic5}).
We make the Fourier transformation for
Eq.~(\ref{eq:Reform}) using the relation,
\begin{equation}
\label{eq:Fourier}
\ww_{\ell,\gamma}(t)=\frac{1}{\sqrt{N}}\sum_{\kk,\gamma}\qq_{\kk\gamma}(t)
\e^{i\kk\cdot \rr_\ell},
\end{equation}
which leads, in the form applicable to both cases of the symmetry-sensitive and the symmetry-insensitive couplings,
\begin{eqnarray}
\label{eq:Fourier_anharmonic5}
\bar{V}_2=\frac{1}{2}\sum_{\kk,\gamma}\bar{\xi}_\gamma\vert \qq_{\kk\gamma}\vert^2.
\end{eqnarray}
Here, for the symmetry-sensitive coupling, the force constant $\bar{\xi}_\gamma$ should be understood as
 $\bar{\xi}_\|\to\tilde{\xi}_s$ and $\bar{\xi}_\perp\to\tilde{\xi}_\theta$ by corresponding stretching motion to longitudinal mode and librational motion to transverse modes
given in Eq.~(\ref{Effective}).
For the symmetry-insensitive coupling,
 the force constant $\bar{\xi}_\gamma$ in Eq.~(\ref{eq:Fourier_anharmonic5}) should be
 independent of $\gamma$ and  distributed from $\tilde{\xi}_\theta$ to $\tilde{\xi}_s$~$(\tilde{\xi}_\theta\le\bar{\xi}_\gamma\le\tilde{\xi}_s)$ as mentioned in the above paragraph.
Thus, we have incorporated the effect of randomly oriented $\UU_\ell$ into the strength distribution of the coupling constants $\bar{\xi}_\gamma$.


The relative displacement $\lbrace\ww_\ell\rbrace$
from the center of the cage induce
dynamic electric-dipoles for both of ONS and OFS as discussed by Fano.~\cite{Fan60}
The inter-site dynamic dipole-dipole interaction $V_3$ is
expressed as
\begin{eqnarray}
\label{eq:dipoles2}
V_3=\sum_{\ell,\ell'}\frac{e_G^{*2}}{2R_{\ell\ell'}^3}[\ww_\ell\cdot\ww_{\ell'}-3(\ww_{\ell}\cdot\hat{\RR}_{\ell\ell'})(\ww_{\ell'}\cdot\hat{\RR}_{\ell\ell'})],
\end{eqnarray}
where the dipole $\pp_\ell$ is defined by
$\pp_\ell=e_G^*\ww_\ell$, and $\hat{\RR}_{\ell\ell'}$ is
a unit vector in the direction to the vector $\rr_{\ell\ell'}=\rr_\ell-\rr_{\ell'}$.
The Fourier transformed expression for Eq.(\ref{eq:dipoles2}) under the random phase approximation
is obtained as
\begin{align}
\nonumber
V_3&=\frac{(e^*)^2}{4}\sum_{\ell,\kk,\gamma}
\vert \qq_{\kk\gamma}\vert^2\mathrm{e}^{-i\kk\cdot\rr_{0\ell}}
\frac{1-3(\hat{\qq}_{\kk\gamma}\cdot\hat{\RR}_{0\ell})^2}
{R_{0\ell}^3}\\
&+ \mathrm{c.c.},
\label{eq:Fourier_Dipoles2}
\end{align}
where $\hat{\qq}_{\kk\gamma}$ means the polarization vector of the mode $\gamma$,
and $\textrm{c.c.}$ indicates complex conjugate of the first term.

With using the Lorentz sums, Eq.~(\ref{eq:Fourier_Dipoles2}) can be expressed as
\begin{eqnarray}
\label{eq:Fourier_Dipoles4}
V_3=\frac{m\omega_p^2}{2}\sum_{\kk\gamma}\vert \qq_{\kk\gamma}\vert^2L_\gamma,
\end{eqnarray}
where $L_\gamma=(\hat{\qq}_{\kk\gamma}\cdot\hat{\kk})^2-\frac{1}{3}$
for small $\kk$  of cubic lattice and
$\omega_p^2=4\pi n_B(e^{*})^2/m$ is the squared plasma frequency of rattling guest ions with mass $m$ and charge $e_G^*$ in cgs esu units.
$n_B=1/a^3$ is its number density charge.
We should note that $L_\|=2/3$ for longitudinal mode
and $L_\perp=-1/3$ for transverse modes, respectively~\cite{Coh55}.

The total kinetic energy of the cages ($M\sum_{\ell,\gamma}\vert\dot{\rr}_{\ell,\gamma}\vert^2/2$) and rattling guest ions ($m\sum_{\ell,\gamma}(\dot{\UU}_{\ell,\gamma})^2/2$)
is given by
\begin{align}
\label{eq:kinetic}
\nonumber
K
&=\frac{m+M}{2}\sum_{\kk,\gamma}\vert \dot{\QQ}_{\kk\gamma}\vert^2
+\frac{m}{2}\sum_{\kk,\gamma}\vert \dot{\qq}_{\kk\gamma}\vert^2\\
&+ \frac{m}{2}\sum_{\kk,\gamma}\left( \dot{\QQ}_{\kk\gamma}\cdot \dot{\QQ}_{\kk\gamma}^*+\dot{\QQ}_{\kk\gamma}^*\cdot \dot{\QQ}_{\kk\gamma}\right),
\end{align}
where the Fourier transformation $\ww_\ell \to \QQ_{\kk}$ is made
by using Eq.~(\ref{eq:Fourier}).


The equations of motion for two variables $\qq_{\kk\gamma}, \QQ_{\kk\gamma}$
with $\gamma=\vert,\perp$ are obtained from the Euler-Lagrange equation as
\begin{gather}
\label{eq:motion3}
m\omega_{\kk\gamma}^2(\qq_{\kk\gamma}+\QQ_{\kk\gamma})
=(\bar{\xi}_\gamma+m\omega_p^2L_\gamma)\qq_{\kk\gamma},\\
\label{eq:motion4}
(m+M)\omega_{\kk\gamma}^2 \QQ_{\kk\gamma}=-m\omega_{\kk\gamma}^2\qq_{\kk\gamma}
+4f_{\gamma}\sin^2\left( \frac{\kk\cdot\aa}{2}\right) \QQ_{\kk\gamma}.
\end{gather}
Here, the above equations of motion is given of the form applicable to both cases of
symmetry-sensitive and symmetry-insensitive couplings.
It is straightforward, for example, to extend these equations to the symmetry-insensitive case by replacing $\bar{\xi}_\gamma\to\bar{\xi}$ ($\tilde{\xi}_\ell\leq\bar{\xi}\leq\tilde{\xi}_s$).

\subsection{On-Center System}
For ONS, we impose the condition
$\xi>0$, $\eta>0$ in Eq. (\ref{eq:anharmonic}).
This yields $\UU_\ell (t)=0$.
For this situation, we have the Fourier transformed $V_2$ as
\begin{eqnarray}
\label{eq:coffee}
V_2=\frac{1}{2}\sum_{\kk,\gamma}\hat{\xi}\vert \qq_{\kk\gamma}(t)\vert^2,
\end{eqnarray}
where $\hat{\xi}=
\xi+\eta\langle\vert\ww_{\ell}(t)\vert^2\rangle/2$.
The characteristics of the force constants $\tilde{\xi}_\gamma$ in Eq.~(\ref{eq:Fourier_anharmonic5}) for OFS and $\hat{\xi}$
in Eq.~(\ref{eq:coffee}) for ONS
are quite different.
The crucial difference is that $\tilde{\xi}_\gamma$ for OFS involves both the librational freedom ($\tilde{\xi}_\theta$)
and the radial one ($\tilde{\xi}_s$), while $\hat{\xi}$ for ONS is isotropic between longitudinal and transverse modes.
The electrostatic interaction $V_3$ for ONS becomes the same as
Eq.~(\ref{eq:Fourier_Dipoles4}).
The equations of motion for two variables $\qq_{\kk\gamma}, \QQ_{\kk\gamma}$are obtained by replacing the parameter as $\bar{\xi}_\gamma\to\hat{\xi}$ in Eq. (\ref{eq:motion3}).


\section{Phonon Dispersion Relations for OFS and ONS}
We determine
the force constants $f_\gamma$
and  $\tilde{\xi}_\gamma$ in Eqs.~(\ref{eq:motion3}) and (\ref{eq:motion4}) from the velocities of sounds and the data of optic spectroscopy.
The force constants $f_\|,f_\perp$ can be
obtained from the sound velocities  $v_\|=3,369$~m/sec and $v_\perp=1,936$~m/sec for $\beta$-BGS ($\beta$-Ba$_8$Ga$_{16}$Sn$_{30}$) of OFS~\cite{Ish06} using the relation $v_\gamma=a\sqrt{f_\gamma/(m+M)}$.
Thus, we have $f_\|=18.1$~N/m and $f_\perp=5.77$~N/m, employing masses $m=136$~u for Ba atom and $M=8.55m$ fro cage, and $a=11.68$~\AA.
The force constants $f_\|,f_\perp$ for $\beta$-BGG ($\beta$-Ba$_8$Ga$_{16}$Ge$_{30}$) of ONS can be obtained using the sound velocities
$v_\|=4,096$~m/sec and $v_\perp=2,795$~m/sec as well~\cite{Chr08}.
These yield $f_\|=26.4$~N/m and $f_\perp=12.3$~N/m with using $M=7.04m$ and $a=10.78$~\AA.

The force constants $\tilde{\xi}_\gamma$ in Eq. (\ref{eq:Fourier_anharmonic5}) and $\hat{\xi}$ in Eq. (\ref{eq:coffee}) can be determined from the data of optic spectroscopy since the spectra for optic modes provide the information at $\kk$=0,  namely, $\Gamma$-point.
Taking $\kk\to 0$ in Eq. (\ref{eq:motion4}) and combining it with Eq.(\ref{eq:motion3}), we have the squared eigenfrequency for optic modes given by
\begin{eqnarray}
\label{eq:optic_eigenfrequency}
\omega_{0\gamma}^2=\frac{\bar{\xi}_\gamma+m\omega_p^2L_\gamma}{m}\left( 1+\frac{m}{M}\right),
\end{eqnarray}
where $\gamma$ denotes the longitudinal or transverse polarization of optic modes.
Equation (\ref{eq:optic_eigenfrequency}) involves the squared plasma frequency $\omega_p^2$ arising from the fluctuation of charged rattling guest ions.
We can estimate the magnitude of the plasma frequency to be $\omega_p/2\pi\approx 0.09$~THz using mass ($m=137$~u) and the charge of Ba$^{2+}$ ($e_G^*=2e$)
by taking the number density $n_B=0.628\times10^{27}$~m$^{-3}$ and the relative electric susceptibility $\epsilon_r/\epsilon_0\approx 10$~\cite{Nak08}, the same one as that of Si crystal.

We can estimate, using the expression Eq.~(\ref{eq:optic_eigenfrequency}), the force constant $\tilde{\xi}_\gamma$ for $\beta$-BGS of OFS
employing the data of Raman scattering~\cite{Tak06} and far infrared spectroscopy~\cite{Mor09}.
Mori \textit{et~al.} ~\cite{Mor09} have recently observed infrared active spectra at 7~K with the lowest-lying peak at
$0.71$ THz with line-width broadening of 0.57~THz
for $\beta$-BGS of OFS by means of THz time-domain spectroscopy.
The peculiar line broadening with decreasing temperature found by Mori $et~al.$ ~\cite{Mor09} is interpreted as the effect of random orientation of dipoles as mentioned in Sec. IV.

The spectrum of A$_{1g}$ stretching mode coupled with longitudinal mode is not observed for $\beta$-BGS of OFS due to technical reasons ~\cite{Tak06}.
For other type of OFS, e.g., $\beta$-SGG, the spectra at 2~K are obtained at 48 cm$^{-1}$,
and the spectrum at 36 cm$^{-1}$ for $\beta$-EGG of OFS for A$_{1g}$ mode ~\cite{Tak06}.
We employ these to estimate the eigenfrequency of A$_{1g}$ mode
of $\beta$-BGS as
$\omega_{0}/2\pi=30$~cm$^{-1}$ ($=0.9$ THz), which we use to evaluate the force constant $\tilde{\xi}_\|$ relevant to longitudinal optic mode.

Under these assignments, we estimate $\tilde{\xi}_\|=7.32(1\pm 0.25)$~N/m and $\tilde{\xi}_\perp=2.21(1\pm 0.25)$~N/m,
using the estimated plasma frequency $\omega_{p}/2\pi$ = 0.09 THz for Eq.~(\ref{eq:optic_eigenfrequency}).
The contribution to optic eigenfrequencies from this plasma frequency is only 10\% at $\Gamma$-point.
The force constant $\hat{\xi}$ for $n$-type $\beta$-BGG of ONS is obtained  at 2 K in a same manner using the data of Raman scattering $\omega_{0}/2\pi=32$~cm$^{-1}$ ($=0.96$~THz)
~\cite{Tak06} as $\hat{\xi}=8.23$~N/m.

\begin{figure}[t]
\begin{center}
\includegraphics[width = 0.25\linewidth]{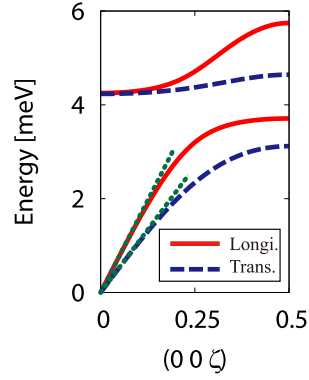}
\caption{(Color online)~Phonon dispersion curves along the [001] direction for $\beta$-BGG of ONS.
Dispersion relations for longitudinal modes are plotted with solid lines and for transverse modes with dashed lines.
Dotted linear lines from the origin represent the long-wave limit of acoustic dispersion relations.
The degeneracy of optic modes at $\Gamma$-point is observed at 4.2 meV arising from the on-cetered symmetric potential.
}
\label{fig:dispersion_BGG}
\end{center}
\end{figure}

We show, at first, the calculated dispersion relations in
 Fig.~\ref{fig:dispersion_BGG} for  $\beta$-BGG of ONS using the force constants obtained above
for Eqs.~(\ref{eq:motion3}) and (\ref{eq:motion4}).
The acoustic phonon dispersions in Fig.~\ref{fig:dispersion_BGG} for $\beta$-BGG
are flattened below $\omega_{0\gamma}$ for both transverse and longitudinal modes.
We emphasize that the eigenfrequencies of optic modes at $\Gamma$-point are degenerate.
The experimental data on the dispersion relations in terms of coherent inelastic neutron scattering are available for $n$-type $\beta$-BGG of ONS~\cite{Lee07, Chr08}.
Our calculated results shown in Fig.~\ref{fig:dispersion_BGG} are good agreement with
the inelastic neutron scattering  data for $\beta$-BGG of ONS~\cite{Lee07, Chr08}.

\begin{figure}[t]
\begin{center}
\includegraphics[width = 0.5\linewidth]{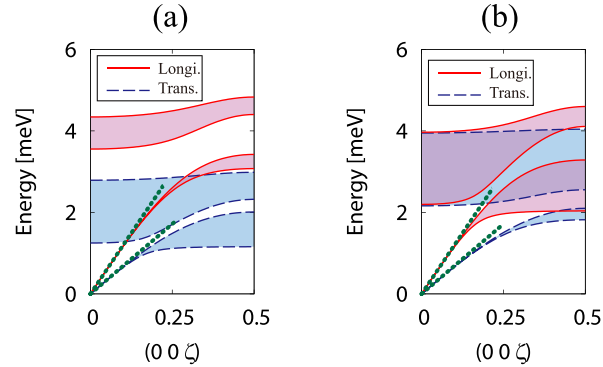}
\caption{
(Color online)~Phonon dispersion curves along the [001] direction for $\beta$-BGS of OFS.
The figure (a) represents the symmetry-sensitive coupling;(b), the symmetry-insensitive coupling.
Band dispersion relations for longitudinal (transverse) modes are
illustrated by the region bounded by solid (dashed) lines in both of (a) and (b).
Dotted linear lines from the origin represent the long-wave limit of acoustic dispersion relations.
Note that optic modes observed at $\Gamma$-point are separated at 2.2 meV and 4.1 meV in the case (a), but not for the case (b).
}
\label{fig:dispersion_BGS}
\end{center}
\end{figure}
Next, we present the calculated dispersion relations for $\beta$-BGS of OFS for both  of symmetry-sensitive and symmetry-insensitive couplings in Fig.~\ref{fig:dispersion_BGS} (a) and (b).
As described in Sec. IV, these two couplings provide qualitatively different line-width broadenings.

It is remarkable in Fig. \ref{fig:dispersion_BGS} (b) that the spectral width for symmetry-insensitive coupling becomes
broader and gapless in the region of avoided crossing attributing to the coupling between acoustic modes and guest ions.
While, the spectral broadening of the symmetry-sensitive coupling shown in Fig.~\ref{fig:dispersion_BGS} (a) reflects the width due to
dipole-dipole interaction between randomly oriented dipoles.
In order to determine which coupling is dominant, the experimental information of inelastic neutron scattering (INS) is required.

Quite recently, INS measurements for both $\beta$-BGS of OFS and $\beta$-BGG of ONS were performed in the temperature range from 5~K to 290~K using the cold neutron disk-chopper spectrometer AMATERAS installed in Materials and Life Science Experimental
Facility (MLF), Japan Proton Accelerator
Research Complex (J-PARC)~\cite{Nak10}.
Preliminary results for $\beta$-BGS at 5~K indicate that the line-width broadening for the lowest band is about 0.5~THz around the main peak of 0.53~THz.
There is no clear gap in the density of states between optic and acoustic branches,
which is distinctly different from the results for $\beta$-BGG.
These suggest that both coupling mechanisms are working.

\section{Thermal conductivities for OFS}

The frequency width
 at the avoided crossing is obtained using Eqs.~(\ref{eq:motion3}) and (\ref{eq:motion4}) as
\begin{eqnarray}
\label{eq:anti_eigenfrequency}
\delta\omega_{c\gamma}\cong\left[\frac{m}{M}\left(\frac{\bar{\xi}_\gamma}{m}+\omega_p^2L_\gamma\right)\right]^{1/2}
\cong\omega_{0\gamma}\sqrt{\frac{m}{M}},
\end{eqnarray}
where the last relation holds for a small contribution from the plasma frequency.
This relation is valid for in our argument since the contribution from the plasma frequency is only 10\% of $\omega_{0{\perp}}$ for $\beta$-BGS.
Equation (\ref{eq:anti_eigenfrequency}) indicates that the
frequency $\delta\omega_{c\gamma}$ at avoided crossing is governed
by the quantities $\omega_{0\gamma}$ and the square root of mass ratio $\sqrt{m/M}$.
The
frequency
 $\delta\omega_{c\gamma}$ for OFS is much smaller than the case for ONS due to the inequality $\bar{\xi}_\gamma<\hat{\xi}$.

The formula of phonon thermal conductivity $\kappa_{ph}(T)$ is given by
\begin{equation}
\label{eq:thermal_cond}
\kappa_{ph}(T)=\frac{1}{3}\sum_{\gamma}\int_0^{\omega_{c\gamma}} C_{ph,\gamma}(\omega)v_\gamma\ell_\gamma\mathrm{d}\omega,
\end{equation}
where $C_{ph,\gamma}$, $v_{\gamma}$, and $\ell_{\gamma}$ are the specific heat, group velocity, and mean-free path of acoustic phonons of the mode $\gamma$, respectively.
The cut-off angular frequency $\omega_{c\gamma}$ of the integral in Eq.~(\ref{eq:thermal_cond}) represents the cross-over frequency from extended phonons to quasi-localized phonons of the mode $\gamma$.
For both of ONS and OFS, $\omega_{c\gamma}$ corresponds to the frequency at the avoided crossing since
the eigenmodes of acoustic branches at the avoided crossing are strongly hybridized with the optic modes originating from rattling guest ions.
Thus, we need to discuss the contribution from extended acoustic phonons below $\omega_{c\gamma}$.

Figure \ref{fig:dispersion_BGS} (a) for $\beta$-BGG of ONS shows that the avoided crossings occur at $\vert\kk\vert>\vert\GG\vert/4$, which allows the Umklapp process $\kk_1+\kk_2=\kk_3+\GG$ for acoustic modes consisting of elastic vibrations of cages.
While, the avoided crossing of transverse acoustic modes, mainly contributing to phonon heat transport,  for $\beta$-BGS of OFS occurs at $\vert\kk\vert\approx\vert\GG\vert/4$ as given in
Fig.~\ref{fig:dispersion_BGS} (b), indicating that acoustic phonons carrying heat are limited to those with the wave number $\vert\kk\vert<\vert\GG\vert/4$.


The cross-over frequency $\omega_{c\gamma}$ is obtained by subtracting the the frequency $\delta\omega_{c\gamma}$ given by Eq. (\ref{eq:anti_eigenfrequency}) as $\omega_{c\gamma}=\omega_{0\gamma}-\delta\omega_{c\gamma}$.
This leads to
\begin{equation}
\label{eq:avoided_freq}
\omega_{c\gamma}\simeq
 \omega_{0\gamma}\left( 1-\sqrt{\frac{m}{M}}\right).
\end{equation}
Since the peak frequency of thermal distribution at the temperature $T$ for extended acoustic phonons becomes $\hbar\omega_p\simeq 3.8k_B T$ taking into account the Stefan shift,
our analysis indicates that the relation holds between the lowest optic eigenfrequency $\omega_{0\perp}$ and the onset temperature $T_p$ of the plateau
expressed as, from Eq. (\ref{eq:avoided_freq})
\begin{equation}
\label{eq:crit_freq}
T_p\approx\hbar\omega_{0\perp}\frac{\left( 1-\sqrt{m/M}\right)}{3.8k_B}.
\end{equation}
This is a general rule applicable to other types of type-I clathrates with off-center rattling guest ions.
\section{Conclusions}
We have investigated the THz-frequency dynamics of type-I clathrates by highlighting the differences between the on- and off-center systems.
Concerning the low-lying optic mode, we have shown the energy range of the transverse mode in $\beta$-BGS of OFS is much lower than that in $\beta$-BGG of ONS, whose origin is due to the librational motion of off-center rattling guest ions.

As for the dispersion relations, our calculated results shown in Fig.~\ref{fig:dispersion_BGG} are good agreement with the experimental data for $\beta$-BGG~\cite{Lee07, Chr08} in terms of inelastic neutron scattering.
The dispersion relations for $\beta$-BGS given in Fig.~\ref{fig:dispersion_BGS} (a) and (b)
show a wide flat region for acoustic phonon branch of transverse modes arising from the low-lying librational optic mode.
About the heat transport, we have found that the distinct difference of  thermal conductivities between $\beta$-BGG and $\beta$-BGS
is attributed to the position of avoided crossing in the Brillouin zone, where the hybridization between optic modes and acoustic modes occurs.

To conclude, our model analysis serves as gaining deeper insight into the physical origins of peculiar low-lying THz-frequency dynamics observed for type-I clathrates with rattling guest ions.
The formula Eq. (\ref{eq:crit_freq}) between $\omega_{0\perp}$ and $T_p$ can used as a rule for designing efficient thermoelectric materials with suppressed glass-like thermal conductivities.
In addition, we claim that the physical mechanism proposed in this paper is applicable for elucidating the origin of the boson peak and plateau thermal conductivity observed in glasses~\cite{Nak02} or relaxers~\cite{Tac09}.

\begin{center}
\textbf{ACKNOWLEDGEMENTS}
\end{center}

We are grateful to  M. Nakamura and M. Arai for providing us neutron scattering data prior to publication.
We thanks C. H. Lee, T. Mori, T. Takabatake, Y. Takasu, and N. Toyota for stimulating discussions.
This work was supported in part by the Grand-in-Aid for Scientific Research(C) and on Priority Areas of new Materials Science using Regulated Nano-Spaces from the Ministry of Education, Culture, Sports, Science and Technology (MEXT) of Japan.
E. K. acknowledges a support from Yukawa Memorial Foundation and the Grand-in Aid for Scientific Research for Young Scientists(B) from MEXT.
A part of the numerical calculation in this work was carried out at the Yukawa Institute Computer Facility


\end{document}